# Flow-aware Forwarding in SDN Datacenters Using a Knapsack-PSO-Based Solution

Sahar Abdollahi[1], Arash Deldari[2], Hamid Asadi[3], AhmadReza Montazerolghaem[4], Sayyed Majid Mazinani[5]

*Abstract*—With the rapid growth of different massive applications and parallel flow requests in Data Center Networks (DCNs), today's providers are confronting challenges in flow forwarding decisions. Since Software Defined Networking (SDN) provides fine granular control, it can be intelligently programmed to distinguish between flow requirements. The present article proposes a knapsack model in which the link bandwidth and incoming flows are modeled as a knapsack capacity and items, respectively. Furthermore, each flow consists of two size and value aspects, acquired through flow size extraction and the type of service value assigned by the SDN controller decision. Indeed, the current work splits the incoming flow size range into Type of Service (ToS) decimal value numbers. The lower the flow size category, the higher the value dedicated to the flow. Particle Swarm Optimization (PSO) optimizes the knapsack problem and first forwards the selected-flows by KP-PSO, and the non-selected-flows second. To address the shortcomings of these methods in the event of dense parallel flow detection, the present study puts the link under the threshold of a 70% load by simultaneous requests. Experimental results indicate that the proposed method outperforms Sonum, Hedera, and ECMP in terms of flow completion time, packet loss rate, and goodput regarding flow size requirements.

## I. INTRODUCTION

THE fast growth of applications poses a great challenge to Data Center Networks (DCNs) due to performance demands. Since modern DCNs should be capable of giving a fast response to concurrent application requests [1], Traffic Engineering (TE) and flow-aware methods have recently turned into fundamental solutions.

Due to the recent rapid growth of data center applications, the DCNs concurrent traffic has complicated traditional networks, especially in terms of dynamic and fine-granular flow specific considerations. Therefore, due to the necessity of acquiring a global view of the network, Software Defined Networking (SDN) has emerged as a widely adopted paradigm in DCNs. Decoupling the control plane and data plane is a fundamental principle in SDN architecture design [1]–[3]. A multitude of researchers and industries intend to achieve better performance parameters through SDN's specific characteristics, such as network programmability, open interfaces, centralized administration, and abstraction, all of which, in turn, lead to agility and flexibility [4]–[7]. Thus, implementing DCN scenarios on SDN offers the proper solutions that result in intelligent decisions.

Owing to the ever-increasing number of cloud service users, DCNs have become a fundamental infrastructure for hosting different applications, such as web services, big data analytics, and cloud applications [8]–[10]. Moreover, current DCNs (e.g., Google cloud and Amazon) should meet the Quality of Service (QoS) requirements. Accordingly, bandwidth and latency ought to be considered as two of the most precious and determinative parameters across servers [11], [12].

The pattern of frequently generated traffic and the amount of unpredictable dynamic traffic in DCNs can impose an unmanageable workload on the network, which may degrade performance and QoS parameters. Additionally, the occurrence of phenomena may be inevitable in some cases and result in link under-utilization, such as the case with TCP incast stemming from uncontrolled TCP behavior and a many-to-one traffic paradigm [13]. Moreover, due to the high cost of ternary content-addressable memory modules, the number of predefined flow entries should be limited [14].

DCNs adopt various approaches such as: high concurrent flow transmission, reactive and proactive TE methods, and suppressing traffic congestion [15]. Although each method pinpoints a particular problem, they may somehow neglect fundamental issues like underutilization, overutilization, and computational overhead. Indeed, DCNs should be capable of providing both delay-sensitive and throughput-sensitive application demands.

Much research [2], [10], [13], [16]–[21] has investigated DCN traffic and related flow management challenges either in SDN or other networking paradigms. Most of the proposed solutions in the DCN context take advantage of leveraging the SDN centralized control and the path diversity typically available in such networks.

Taking typical DCN flow types into consideration can reduce the shortcomings of current traffic management methods. Statistical studies conducted in DCNs [10], [22], [23] specify two typical flow types, namely Mice Flows (MFs) and Elephant Flows (EFs). MFs comprise 80% of the total flows in DCNs (e.g., emails and web pages) and are less than 100 KB in size. Also, MFs are short-lived and delay-sensitive (lasting less than a few milliseconds). On the other hand, EFs make up 10% of the total flows in DCNs (e.g., VM migrations and MapReduce) and are more than 100 KB in size. EFs are also long-lived (lasting minutes or more) and throughput-sensitive [10], [23], [24].

[1] Department of Computer Engineering, Salman Institute of Higher Education, Mashhad, Iran. Email: sa.abdollahi@salman.ac.ir
[2] Department of Computer Engineering, University of Torbat Heydarieh, Torbat Heydarieh, Iran. Email: adeldari@torbath.ac.ir
[3] Department of Electronic Engineering, Islamic Azad University of Islamshahr, Islamshahr, Iran. Email: hamid93asd@gmail.com
[4] Faculty of Computer Engineering, University of Isfahan, Isfahan, Iran. E-mail: a.montazerolghaem@comp.ui.ac.ir
[5] Department of Computer Engineering, Imam Reza International University, Mashhad, Iran. Email: smajidmazinani@imamreza.ac.ir

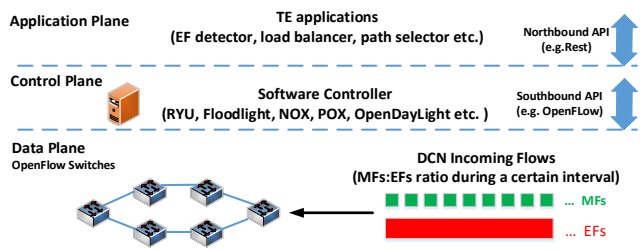

Fig. 1. MFs and EFs in SDN datacenters

A concentration of EFs in DCNs can lead to a load imbalance because the Equal-Cost Multi-Path (ECMP) routing maps them to the same path. As a result, the load imbalance affects other flows attending for the utilization of the shared link [10]. To propose forwarding strategies for different flow types, some researchers concentrate on flow detection, particularly for EFs in an SDN environment [22]. However, they might neglect MFs and the imposed overhead of the EF detection. To prevent MFs from becoming stuck between EFs, which leads to a load imbalance, DC flow sizes should be considered.

As a solution to address the issues of flow size and its requirements in SDN DCNs, the present paper introduces Size-KP-PSO which is comprised of a knapsack model optimized with the particle swarm optimization algorithm that considers DCNs flow sizes to perform TE. A flow tagging approach is proposed that deals with concurrent parallel incoming flows while making distinctions between their requirements. To the best of the current study's knowledge, this is the first flow-aware forwarding modeled through the network link with the knapsack problem optimized by the PSO algorithm.

The main contributions of the present article are as follows:
- Tackling the problem of concurrent flow requests through an integrated handling of flow size requirements for both MFs and EFs.
- Proposing the Size-KP-PSO method which arranges flow size tag in a spectral range rather than dividing it into only two specific sizes (MFs and EFs).
- Dedicating two features of volume (weight) and value (significance) to each flow so as to organize the classification of flow size requirements.
- Leveraging our customized type of service (ToS) as the flow significance (value) and then injecting it into the knapsack value vector.
- Modeling the problem as a knapsack problem that aims to fill the link capacity through leveraging flow size while picking the most valuable incoming flow requests (selected-flows) first.
- Sending the remaining flows afterwards which mostly contains EFs (non-selected-flows).
- Overcoming the bottlenecks of MFs while improving the satisfaction of EFs through integrated flow size scheduling.
- Handling concurrent flow requests by utilizing multithreading and parallel flow forwarding policies besides flow scheduling.

The rest of this paper is organized as follows. Section II presents a literature review on traffic engineering in SDN DCNs. Section III introduces the proposed method while Section IV describes the evaluations in detail. Section V provides the conclusion and discusses future research directions.

## II. RELATED WORKS

### A. SDN background

The SDN architecture provides abstraction through the separation of the data plane and control plane, as illustrated in Fig. 1. The key feature of the SDN paradigm is programmability, which is enabled through a centralized control plane [22]. This characteristic allows service providers to develop their applications and satisfy intra-data center and user demands. Moreover, because a majority of tasks, except forwarding, are delegated to the controller, the data plane can be composed of cost-effective commodity switches. The SDN controller communicates with the data plane and application layer through south-bound and north-bound APIs, respectively.

The provided interaction between switches and the controller entity greatly simplifies intelligent decision making for the network. For instance, when an OpenFlow switch receives a packet on a port, it looks up the flow table to match the packet to a flow entry rather than working independently of the rest of the network. If no match is found, the OpenFlow switch sends a packet-in to the controller for further processing [25].

The SDN controller acts as a network operating system and prepares a global view of the network state. Hence, various tasks, such as TE functions, load balancing strategies, and EF detection, have already been implemented by SDN and have received significant attention in recent research [26].

### B. Traffic engineering without considering flow size

Traffic engineering and flow scheduling research mostly do not inspect flows according to their size, but concentrate on the prediction and prevention of congestion and routing.

Pereira et al. [27] propose TE with segment routing that utilizes three segments for each SR path configuration. The method enables SDN controllers to perform optimized SR configurations. SALP-SR achieves near-optimal network resource utilization based on the interior gateway protocol configuration and by computing the traffic amount to be forwarded along with neighboring segments. The computational overhead, however, should be considered.

Sacco et al. [28] introduce RoPE as an architecture that performs a bandwidth prediction method with SDN controller cooperation. RoPE adapts the routing strategy and takes a new route based on congestion prediction. Although RoPE overcomes edge-based application problems, it still might not fit all use cases.

Ren et al. [1] propose a TE method through integer linear programming formulation. This method redirects each flow to the proper node(s) and then splits them into multiple paths. Although this minimizes the maximum utilization of the link, it functions in static traffic conditions.

TABLE I
EF DETECTION CATEGORIES[15], [13], [22]

| Categories | Methods | Description | Hardware Changes | Overhead | Cost |
|---|---|---|---|---|---|
| Detecting Location | Switch-based | Whole or partial detection is performed on the switches. | Yes | In proportion to accuracy | In proportion to accuracy |
| | Host-based | The detection burden shifts to end hosts. | Yes | Low | Low |
| Traffic Information Collection Technique | Active Measurement | The SDN controller sends Read-State messages periodically to the switches and then receives feedback | No | High | High |
| | Passive Statistics | If no flow entries match the new packet, both the switch and controller initiate interactions. | Yes | Low | Low |
| | Sampling | It mandates switch capture packets at a certain rate. | Yes | Medium | In proportion to accuracy |

In [16], Luo et al. introduce a preemptive multicast scheduling approach for DCNs. The method utilizes a simple single hop to split multicast, which leads to a reduction in flow completion time. Ruan et al. [29] propose a priority-based transport control protocol which has a recovery approach through packet priority while also tackling congestion. In addition, the protocol contemplates queue length at the bottleneck switch.

Shaffan et al. [30] propose a traffic management approach that is conducted according to the prioritization of flow media types (also known as Multi-purpose Internet Mail Extensions). This includes text, image, audio, and video through big data in the SDN network. Although this approach maximizes utilization, more parameters should still be considered.

Utilized in most current data centers [24], ECMP [31] provides a strategy allowing the traffic of the same destination to be split through multiple next-hops with equal costs and utilizing packet header hashing [8]. However, since ECMP does not consider the network load, it is not capable of adapting its splitting strategy to dynamic traffic conditions [32]. Moreover, as ECMP maps traffic to the same path, this leads to the concentration of EFs, which subsequently results in packet retransmission and higher latency [10].

Paliwal and Shrimankar [33] propose an energy-saving switch-ON/OFF procedure as well as a rerouting approach that works on SDN-MPLS in data center networks. Although an energy-saving method, this procedure shows improvements in bandwidth utilization and also prevents active traffic loss.

In summary, although improving some aspects of traffic engineering, each of these studies suffer from a lack of a multi-purpose view of flow nature and its requirements. Therefore, rather than dynamically analyzing flows before forwarding them, these approaches confront traffic without any basic information about the flows. The current research utilizes Size-KP-PSO, modeled with the knapsack problem optimized by the PSO algorithm. Flow size is considered before forwarding and an effort is made to satisfy the corresponding requirements.

*C. Traffic engineering considering flow size*

There have been several studies that consider flow size or EF detection for overcoming MF bottlenecks. Some methods directly pinpoint the flow nature and adjust their approach according to mice or elephant flow types.

Modi and Swain [34] propose a dynamic flow scheduling algorithm that figures out the host and link capacity. Whenever a flow request comes in, a module compares the current load status of the nodes and their thresholds. Thereafter, the flow is sent as usual if the load is less than a threshold; otherwise, the flow is assigned to a new path. This method performs better only when dealing with MFs.

Kanagevlu and Aung [35] introduce a local re-routing approach that detects EFs and then reroutes them in the event of congestion or a hop before. This method, however, still has some shortcomings, such as congestion manifestation in the Top-of-Rack (ToR) switches where no other choices are available to carry out rerouting. Liu et al. [36] propose a load balancing mechanism for EFs which first detects EFs and then splits them using a weighted routing algorithm. This method can improve network utilization, but assembling the EFs together leads to an increase in overhead.

*1) EF detection*

Researchers categorize EF detection techniques according to two main aspects, as presented in Table I.

Tan et al. [13] propose Sonum as a software-defined synergetic sampling approach detecting EFs by consolidating EF sampling information from multiple switches and then scheduling the EFs. Although outperforming ECMP and Hedera in terms of Flow Completion Time (FCT) and throughput, this method does not consider link utilization and MF requirements.

Hamdan et al. [22] present an SDN-based flow-aware EF detection which applies two classifiers each on SDN switches and the controller, respectively. The technique enables EF classification task sharing between switches and the controller. It also filters most MFs in the switches which, in turn, reduces request messages to the controller. However, as the switch-side classifier detects more EFs, the amount of false detection of some MFs rises.

Hedera proposes a proactive polling method that detects EFs at edge switches and assigns them to non-conflicting paths [37], [34]. Hedera is a priority-based demand estimation which transfers flows using ECMP forwarding by default. This approach utilizes a central scheduler that computes source-destination bandwidth and utilizes the information for assigning other flows to paths as well [38]. However, because Hedera must repeatedly generate the demand matrix, it imposes a major overhead on the system [37]. Moreover, this method does not consider that immediate MF requests may also lead to network congestion [13].

## III. PROPOSED METHOD

The present study introduces Size-KP-PSO, a novel method that integrates flow size consideration in both EFs and MFs for traffic engineering in the event of dense parallel flows. The proposed approach exploits controller capability that can be intelligently programmed.

The overview of the Size-KP-PSO method is comprised of multi-steps: (1) the initial model components (flow size, flow tag, link bandwidth) are mapped to the knapsack problem's input parameters (weight, value, total knapsack capacity), respectively; (2) the size (weight) of incoming parallel detected flows are extracted through the SDN controller; (3) the SDN controller also determines the flow tag (value) according to flow size; (4) both the flow size and flow tag, in addition to the link bandwidth of next-hop (total knapsack capacity), are sent to the knapsack model; (5) the knapsack problem is optimized through the PSO algorithm to find the fittest solutions, which are the selected-flows to be primarily sent; (6) the other flows, which are the non-selected-flows, are sent. The remainder of this section provides comprehensive details of Size-KP-PSO for each of these steps.

### A. Motivations of modeling with KP

Current DCNs typically deal with dynamic traffic, where both applications which are FCT-sensitive (e.g., web search comprised of MFs) and bandwidth-sensitive (e.g., MapReduce comprised of EFs) co-exist. Moreover, both bursty and smooth DCN traffic may occur simultaneously in a saturated workload [8]. Whereas most data center TE approaches manage congestion while ignoring fine granular flow requirements, the proposed method designs the TE model while taking possible considerations into account.

The widely studied KP problem [39]–[41] is a well-known discrete combinatorial optimization problem. KP places a finite number of items in a knapsack, of which each item has an associated value and weight. The solution is to gain the highest possible value while not exceeding the knapsack total capacity.

In introducing this model, the current article first discusses the motivations behind it and explains how the mapping of network parameters to KP parameters is performed. KP modeling enables attending to multi-purpose factors that converge to flow size management and meeting the flow requirement. The realized considerations that motivated KP modeling are mentioned as follows:

- **Link utilization:** Since the KP problem contains a total weight in which tries to fulfill, resembling link capacity as the knapsack's total weight provides proper link utilization while preventing the link from exceeding the threshold. It also packs the knapsack with the possible items.
- **Flow size consideration:** Initially having a weight vector in itself, the KP model can uniformly cover the flow size attribute. Therefore, the flow size is resembled as the item weight existing in the KP model.
- **Flow prioritization:** Since the KP model comprises the value vector, it can uniformly assign priority based on a predefined desire. Accordingly, the flow significance is resembled as the item value in the KP model.
- **Interaction between size and priority:** KP model orientation embodies weight and the value vector. The objective function presents a qualified interaction between size and priority.

TABLE II
MODELING THE PROBLEM

| Network Parameter | Corresponding Element in Knapsack |
|---|---|
| Link | Knapsack |
| Flow | Item |
| Bandwidth | Total Weight |
| Flow Size | Item Weight |
| Flow Value | Item Value |

### B. Problem Formulation

#### 1) Knapsack Problem

As Table II depicts, the proposed TE model is adapted with KP. Therefore, there is a link to the container (knapsack) with a defined bandwidth (total weight) in which multiple concurrent flows (items) come in. Each flow has an associated volume or size (item weight) and priority (item value). The current work employs the binary knapsack model which is defined by $n$ items, where each item $i$ is associated with the values of $weight_i$ and $value_i$. Also, taking the value of one by the 0-1 decision variable ($dv_i = 1$) denotes the selection of the item. In contrast, zero ($dv_i = 0$) indicates that the item has not been selected.

The knapsack should only carry less or equal to $tc$. In addition, $f$ aims to return the items that should be packed in the knapsack to maximize the total $value$. As it is shown in Table III, network parameters are mapped to the mentioned corresponding elements in the KP.

The current study presents its fitness function $f$ as Eq. (1) [42], which returns the items (flows) with the highest $value$ that should be packed in the knapsack (link). A set of generated parallel flows is considered as items $i$ where $value$ and $weight$ are dependent vectors of flow tag (value) and flow size (weight), respectively.

$$\begin{aligned} f &= \max \sum_{i=1}^{n} value_i * dv_i \\ s.t &\sum_{i=1}^{n} weight_i * dv_i \leq tc \\ dv_i &\in \{0,1\}, i = 1, \dots, n \end{aligned} \quad (1)$$

#### 2) Particle Swarm Optimization (PSO)

The particle swarm optimization [43] is a population-based approach that belongs to the category of evolutionary and swarm intelligence algorithms. It is also a metaheuristic nature-inspired method for solving problems with the minimum amount of information required [44].

For a given optimization problem, each particle in PSO is considered as a solution. In the PSO model, particles are in the d-dimensional space and are comprised of two vectors as represented in Table III [45],[42].

TABLE III
KNAPSACK PARAMETERS, DESCRIPTIONS, AND INITIALIZATIONS

| Symbol | Descriptions |
|---|---|
| $i$ | Item Index (flow index $i \in n$) |
| $n$ | Number of Items |
| $weight_i$ | Item Weight |
| $value_i$ | Item Value |
| $dv_i$ | Objective Function |
| $f$ | Objective Function |
| $tc$ | Total Knapsack Capacity |

The position vector contains the problem variable values. Also, the number of problem parameters is considered as search space dimensions. Moreover, the velocity vector is utilized to update the particle's position. This vector defines the magnitude and direction of the step size for each dimension and each particle independently. The PSO process is described as follows [46]:

**Step 1- Generating initial swarm:** Position $X_i$ and velocity $V_i$ of each particle are initialized randomly based on the allowable range as Eq. (2) and Eq. (3) respectively [47].

$$x_{ij}(0) = \begin{cases} 0 & rand(0,1) < 0.5 \\ 1 & rand(0,1) \geq 0.5 \end{cases} \quad (2)$$

$$v_{ij}(0) = v_{min} + rand(0,1)(v_{max} - v_{min}) \quad (3)$$

**Step 2- Evaluating the objective function:** In Eq. (4) [47], For each particle, $f(x_i)$ is evaluated, which $Q$ is a penalty factor. The objective function value is affected by violations when exceeding the constraint.

$$f(x_i) = \sum_{j=1}^{n} value_i x_{ij}(t) - Q \left| \min\{0, tc - \sum_{j=1}^{n} weight_j x_{ij}(t)\} \right| \quad (4)$$

**Step 3- Updating the best current position:** As depicted in Eq. (5) [46], If $f_{(x_i)}^t$ is better than its $p^{best}$ in history, then $f_{(x_i)}^t$ is set as the new $p^{best}$ of particle $x_{ij}$.

$$if\ f_{(x_i)}^t \leq f_i^{best}, then\ f_i^{best} = f_{(x_i)}^t\ and\ p_i^{best})^t = x_i^t \quad (5)$$

**Step 4- Updating the best global position:** The best particle with the maximum objective function value is identified here. The best particle is employed for updating the positions of other particles. In other words, particles share their information by using the observed characteristics of the best particle. For each particle, compare the $g^{best}$ with its $f(x_i)$, if $g^{best}$ is better, then the index sign of $g^{best}$ will be set again as Eq. (6) [46]:

$$if\ f_{(x_i)}^t \leq f_g^{best}, then\ f_g^{best}\ and\ g^{best})^t = x_i^t \quad (6)$$

**Step 5- Updating the velocity $v_{id}$ and the position $x_{id}$ for each particle:** For each iteration, the dimensional change formulas are expressed as Eq. (7) and Eq. (8) [46].

$$v_i^{(t+1)} = wv_i^t + r_1^t c_1(p_i^{best} - x_i^t) + r_2^t c_2(g^{best} - x_i^t) \quad (7)$$

$$x_i^{(t+1)} = x_i^t + v_i^{(t+1)} \quad (8)$$

**Step 6- Checking the termination condition:** The *maximum iteration* is considered as a stopping criterion. If it meets the condition, it stops; otherwise, it returns to step 2. Since PSO is the optimization method, it finds the best solution for the KP and also plays a fundamental role in the reduction of computation time. Further details of knapsack-PSO symbols, initializations, and descriptions are described in Table IV. The knapsack's solution is optimized using the PSO approach which returns the selected-results. Fig. 2 illustrates a sample of the Size-KP-PSO traffic trend which is captured by sFlow-RT [48]. As Fig.2 depicts, the selected-results with higher values (mostly MFs) are forwarded first. Then the remaining flows (mostly EFs) will be determined to get forwarded into the link. Algorithm 1 demonstrates the PSO approach in the proposed method.

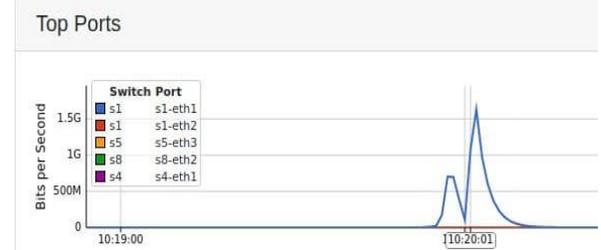

Fig. 2. Size-KP-PSO approach

TABLE IV
PSO PARAMETERS, DESCRIPTIONS, AND INITIALIZATIONS

| Symbol | Description and Initialization |
|---|---|
| $D$ | Number of Dimensions |
| $X_i = (x_{i1}, x_{i2}, ..., x_{iD})$ | $i^{th}$ Particle |
| $V_i = (v_{i1}, v_{i2}, ..., v_{iD})$ | Velocity of $i^{th}$ Particle |
| $C_1, C_2$ | Positive Acceleration Constants, 2 |
| $r_1, r_2$ | Independent Random Values Uniformly Distributed in Range of [0,1] |
| $W$ | Inertia Weight, $\in [0.4, 0.9]$ |
| $Q$ | Sufficiently Large Positive Number |
| $MaxIt$ | Maximum Iterations, 30 |
| $N$ | Number of Particles (population), 300 |
| $V_{min}$ | Minimum Velocity |
| $V_{max}$ | Maximum Velocity |
| $T$ | Increment of Time |
| $x_{ij}^t$ | Position of $i^{th}$ Particle at $t^{th}$ Iteration |
| $v_{ij}^t$ | Velocity of $i^{th}$ Particle at $t^{th}$ Iteration |
| $p_{best}$ | Best Current Position |
| $g_{best}$ | Best Global Position |
| $f(x_i)$ | Fitness Value of $i^{th}$ Particle |
| $f_i^t$ | Current Fitness Value |

**Algorithm 1:** BKP-BPSO [42], [49]
**Input:** N, $MaxIt, Value, Weight, tc$
**Output:** Optimal position $g_{best}, f(g_{best})$

1. while ($t \leq MaxIt$)
2.   $i[1...n] \leftarrow (\{value_i/weight_i|\ value_i \in Value, weight_i \in tc, 1 \leq i \leq n\})$
3.   $for\ i = 1\ to\ N$
4.     $if\ f(X_i) > f(P_i) then\ do$
5.       $for\ d = 1\ to\ D$
6.         $p_{id} = x_{id}$
7.       $end\ for$
8.     $end\ if$
9.     $for\ d = 1\ to\ D$
10.       $apply\ Eq.7$
11.       $v_{id} \in [-V_{max}, V_{max}]$
12.       $apply\ Eq.2$
13.     $end\ for$
14.   $end\ for$
15. $end\ while$
16. $return\ (g_{best}, f(g_{best}))$

## C. Tagging flows by customized ToS

Since flow size plays an initial role in DCNs, the present study proposes a spectral flow tagging approach that distinguishes between the inner sizes of EFs rather than only their unification in a range of more than 100KB. For instance, although a 200MB flow is much heavier than a 150KB flow, and consumes a considerable larger bandwidth, both are typically considered solely as an EF in the DCN flow size classification. Therefore, the flow requirements (e.g., bandwidth and delay) might be neglected. Also, the present study considers the highest tag value for MFs to satisfy their crucial time-sensitivity.

To implement the current work's flow tagging approach, an 8-bit ToS field is utilized in the IPV4 header. In essence, each different DSCP and ECN bit combination is expected to stimulate every network device along the traffic path so that it behaves in a specific way and enable it with a particular QoS treatment for traffic. However, the present study leverages the ToS field potential and designs a customized ToS in order to assign prescribed tags to flows, according to the *ToS tag marking* module implemented on the SDN controller. Hence, the considered tag range is composed of all one octet decimal values regardless of the zero value {1,…,255}.

As shown in Fig. 3, the assumed values are $\{v_i \mid i \in 1,...,255\}$. Furthermore, the flow weight between 1KB and 200MB is considered as $w_i \in \{1KB,...,200MB\}$. Then this bound is split into a ToS vector length equal to 255. Moreover, values (ToS tags) are assigned to weights (flows value) in reverse, which means higher values are assigned to lower weights in that order.

As presented in Algorithm 2 *(lines 10-30)*, the value of 255 is dedicated to MFs (≤100KB) and the other 254 values are divided between the other flow sizes. To be more accurate, regarding the determined ranges of flow size and corresponding ToS value, the lower the flow size, the higher the ToS value will be assigned.

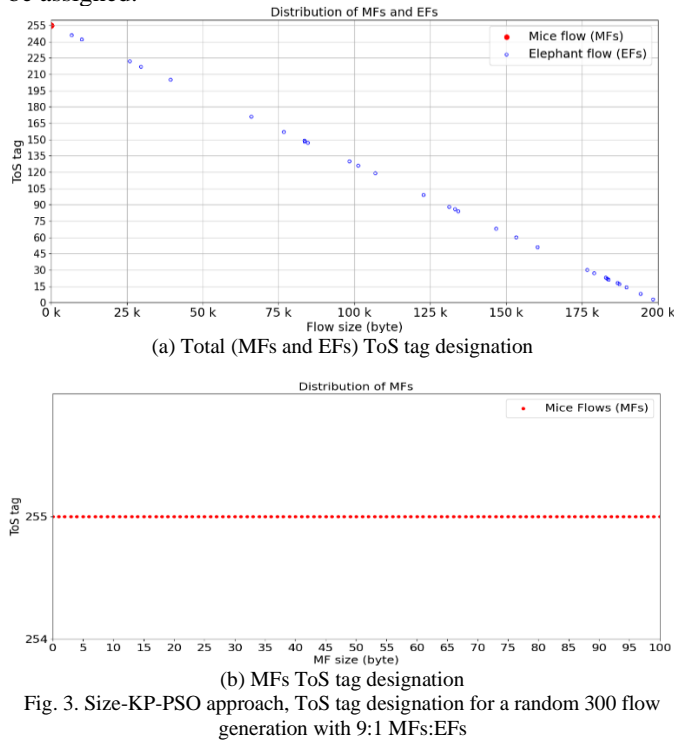

(a) Total (MFs and EFs) ToS tag designation

(b) MFs ToS tag designation

Fig. 3. Size-KP-PSO approach, ToS tag designation for a random 300 flow generation with 9:1 MFs:EFs

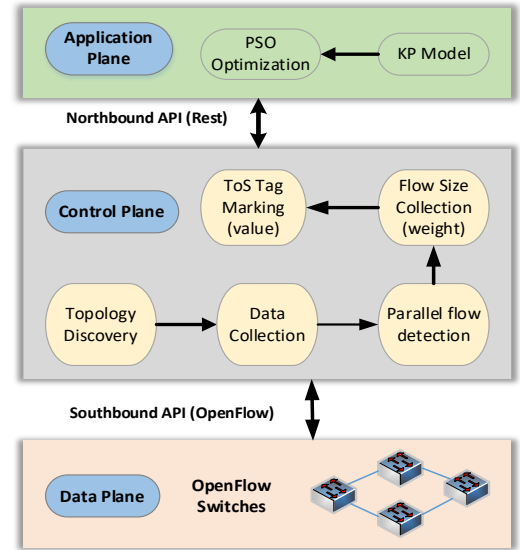

Fig. 4. Size-KP-PSO framework with SDN

## D. Size-KP-PSO Framework with SDN

The overall proposed Size-KP-PSO Framework with SDN is illustrated in Fig. 4. Also, Algorithm 1 represents the process of returning selected-results through the typical form of Binary KP and PSO (BKP-BPSO). Moreover, Algorithm 2 is comprised of whole modules adopted in Size-KP-PSO framework with SDN in which controller modules are presented orderly. In addition, the BKP-BPSO module is evoked in Algorithm 2 to fulfill the process. The details of Algorithm 2 are elaborated as follows:

● *Topology discovery (line 1):* When the controller initiates, it starts sending requests to all switches through *OFPFeaturesRequest* in order to detect connected switches. Afterward, the switches reply by *ofp_event.EventOFPSwitchFeatures* and provide related information to the controller. After establishing initial communication between the controller and switches, the controller eventually begins instructing the switches to send Link Layer Discovery Protocol (LLDP) packets. Then the switches inform the controller about nodes and link statuses by packet-out.

● *Data collection (lines 4-6):* It is necessary to set flow entries on the edge switches in order to send packet-in per-flows even though flows have the same source and destination. To acquire occupied link bandwidth during specified intervals, the module periodically communicates with edge switches through *OFPAggregateStatsRequest*. For an accurate calculation, the switch port statistic, which is connected to the host, is collected. Thus, the bytes crossing the switch are counted in certain intervals through the rest of the API. The link usage is calculated as Eq. (9):

$$LU_t = \frac{BW_{occupied(t)}}{BW_{initial}} \quad (9)$$

where $BW_{occupied(t)}$ and $BW_{initial}$ represent the occupied and initial bandwidth (both in bytes), respectively.

| | |
|---|---|
| **Algorithm 2** :Size-KP-PSO framework with SDN | |
| 1. | Detect topology using LLDP |
| 2. | try |
| 3. |   while True then |
| 4. |     $flows \leftarrow$ all flows in OpenFlow switch |
| 5. |     $BW_{occupied} \leftarrow$ using OFPAggregateStatsRequest |
| 6. |     $UL_t \leftarrow \frac{BW_{occupied}}{BW_{initial}}$ |
| 7. |     $flows_{cm} \leftarrow$ find matching flow rules $(IPv4_{src} ==$ same and $IPv4_{dst} ==$ same and $TCP_{dst_{port}} ==$ same$)$ in flows |
| 8. |     if $UL_t \geq 0.7$ and $flows_{cm}$ then |
| 9. |       use OFPFlowStatsRequest to calculate $flows_{cm}$ weight |
| 10. |       $value \leftarrow sort(range(1 - 255))$ /* ToS tagging */ |
| 11. |       for flow in $flows_{cm}$ do |
| 12. |         if $weight_{cm} \leq 100$ then |
| 13. |           MFs $\leftarrow (255, weight_{cm})$ |
| 14. |         else if $weight_{cm} > 100$ then /* $(200000 - 100)/254 = 787$ */ |
| 15. |           $lower\_bound = 101$ |
| 16. |           $upper\_bound = lower\_bound + 787$ |
| 17. |           $flag = True$ |
| 18. |           while($flag$) do |
| 19. |             if $lower\_bound \leq weight_{cm} \leq upper\_bound$ do |
| 20. |               $division = (upper\_bound)//787$ |
| 21. |               $arg = (value[division], weight_{cm})$ |
| 22. |               EFs $\leftarrow arg$ |
| 23. |               $flag = False$ |
| 24. |             else |
| 25. |               $lower\_bound = upper\_bound + 1$ |
| 26. |               $upper\_bound = lower\_bound + 787$ |
| 27. |             end if |
| 28. |           end while |
| 29. |         end if |
| 30. |       end for /* end of ToS tagging */ |
| 31. |       send MFs and EFs to BKP_BPSO module |
| 32. |       BKP_BPSO $\rightarrow$ return selected |
| 33. |       $non\_selected \leftarrow flows_{cm} - selected$ |
| 34. |       send selected and non_selected to ToR switch /* (which the sender is connected to) */ |
| 35. |       set priority in flow entries for both lists /* (selected has higher priority in comparison to non_selected) */ |
| 36. |       send selected and non_selected respectively |
| 37. |     else |
| 38. |       use simple_switch_stp_13 |
| 39. | except |
| 40. |   use simple_switch_stp_13 |

- *Parallel flow detection (line 7):* When $LU_t$ exceeds the predefined threshold (70% of the initial link bandwidth) while having the same *IPv4_dst* and *TCP_dst* ports, the switch sends a *packet-in* to the controller. The process of checking these conditions is handled by *OFPFlowStatsRequest*. The reply to the request is returned through an event_handler, which is called *ofp_event.EventOFPFlowStatsReply*. If the conditions are true, the next module runs; otherwise, the *simple_switch_stp_13* is executed *(lines 37-40)*.
- *Flow size collection (line 9):* To acquire each flow weight, the byte-count filter of *OFPFlowStatsRequest* is utilized.
- *ToS tag Marking (lines 10-30):* The controller assigns a tag to each flow based on the Size-KP-PSO strategy. Thereafter, it returns the tag vector corresponding to the size vector (weight) of each flow and sends out the tuple (value, weight) to the knapsack model.

- *KP model (line 31):* The module collects the required flow characteristics (flow size (weight), flow tag (value)) for the knapsack model. It also assigns total knapsack capacity through the next-hop initial bandwidth, such that, if the destination is at the edge level (ToR switches), the total knapsack capacity is set to 1GB; otherwise, it is set to 2GB since the next-hop is at the aggregate level anyhow. Afterward, the knapsack model inputs are prepared and ready to be solved by PSO optimization.
- *PSO optimization (lines 32-33):* The optimizer module provides the fittest solution (flows) to the KP. The solution returned to the knapsack model is the items selected (selected-flows) to pack in the knapsack.

Finally, the flow rules on the switches *(lines 34-35)* are installed. The present study installs the selected and non-selected-flows as flow entries on switch flow tables. The *OFPFlowMod*, which allows modification of flow entries, enables designating a higher priority for selected-flows and a lower one for non-selected-flows. Selected-flows are first forwarded to the link, and the non-selected-flows are forwarded second *(line 36)*.

IV. PERFORMANCE EVALUATIONS

This section presents the experimental setup, results, and analytical discussion of the currently proposed methods. For a fair comparison, two typical techniques (Hedera and ECMP) widely investigated in datacenter TE research are chosen. The present study's work is also compared with a recent heuristic long flow scheduling algorithm (Sonum) which has the most similarity to the current work's purposes in terms of flow size consideration, scheduling of detected flows, and a partially identical testbed infrastructure with Size-KP-PSO.

*A. Experimental setup*

The proposed method is implemented using the Mininet network emulator [50] and the Ryu controller [51]. In addition, the details of other experimental setups are presented in the following.

**Topology**: The fat-tree topology is employed as a multi-stage tree-like topology in which switches are interconnected together to constitute three layers: core, aggregation, and edge (ToR) switches. Since the bandwidth of the fat-tree connection increases towards the root of the tree [52], the bandwidth of 1Gbps, 2Gbps, and 4 Gbps are assigned to the layers of the edge, aggregation, and core, respectively. Moreover, the fat-tree is one of the most popular multi-rooted tree topologies [13] which can be deployed to prepare multi-path routing and improve parameter throughput in data center networks [17]. As shown in Fig. 5, the simulation employs a fat-tree topology with 4 pods.

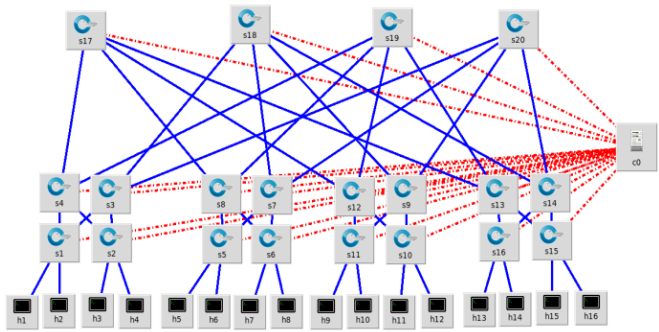

Fig. 5. Fat-tree topology with 4 pods

TABLE V
TRAFFIC PATTERN DESCRIPTION

| Traffic Pattern | Source | Destination |
|---|---|---|
| Stag (p,q) | Any Host | Any Host: In the same edge switch, the same pod and the rest of the network with the probability of *EdgeP*, *PodP*, 1-*EdgeP* – *PodP* respectively. |
| random | Any Host | Any Host |
| Stride (i) | Host (x) | Host (x + i) mod (num hosts) |

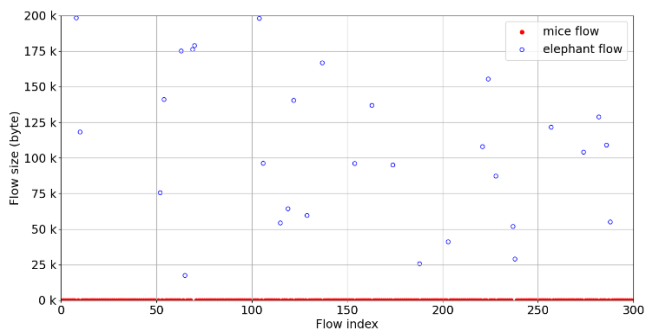

Fig. 6. Distribution of mice and elephant flows

**Traffic patterns:** Similar to Al-Fares et al. [53], the present study generates various traffic patterns as represented in Table V. The main reason for assuming these scenarios is that the over-subscription rate varies across layers. Also, given the fact that the efficiency of the core switches is generally higher and congestion and packet-loss are more likely to occur at the level of the ToR switches, the consideration of these scenarios for sending traffic leads to a fair performance of tests throughout all three layers.

**Traffic generation:** The traffic distribution model is composed of a set of generated flows. As Fig. 6 illustrates, the ratio of 9:1 for MFs: EFs is considered, a common scenario in data center networks. Also, the size range of MFs and EFs varies a bit among different research considerations. However, most frameworks consider the size of MFs and EFs between 1KB-100KB and more than 100K, respectively. Moreover, we applied discrete uniform distribution to generate each MF's and EF's size category.

Also, as presented in Eq. (10), the F set and MF and EF subsets are introduced:

$$F = f_i \mid i = 1, \dots, n\}, n = 300$$
$$F_{mf} = \{f_{mf} \subseteq F \mid mf \leq 100KB\} \quad (10)$$
$$F_{ef} = \{f_{ef} \subseteq F \mid 100KB < ef \leq 200MB\}$$

where $F$, $F_{mf}$, and $F_{ef}$ represent the total set of flow generation, the proportion of MFs, and the proportion of EFs, respectively.

The generated traffic in Iperf is comprised of 300 parallel flows with the aforementioned distribution. Also, the packet headers contain ethernet, IPV4, and TCP fields. Since the maximum standard size of each packet in IPV4 is equal to 65536 bytes, Iperf increases the packet numbers so as to generate larger flows. Moreover, the window size is set on the default value and the packet reordering is adopted.

**Flow forwarding technique:** The current study utilizes a Python multithreading script which is lightweight and requires fewer resources. A thread is dedicated to each flow. In essence, this helps in three ways by (1) enabling accurate simulation of the incoming concurrent (parallel) flows, (2) handling the operation of forwarding all the parallel flows, and (3) reducing the experiments' time and the number of needed processing resources. This technique is applied to all of the methods and scenarios so as to fairly compare the experiments.

**System parameters:** The experiments are conducted on a virtual machine with a quad-core processor and 4 GB RAM. The Ubuntu 18.04 operating system is also used. The detailed simulation parameters can be found in Table VI.

**Methodology:** To demonstrate the performance of the proposed method, Size-KP-PSO is compared with the following state-of-the-art schemes by using TCP as the default transport protocol:

TABLE VI
SIMULATION PARAMETERS

| Parameter | Value |
|---|---|
| Network Topology | Fat-tree |
| Number of Switches | 20 |
| Number of Hosts | 16 |
| Number of Controllers | 1 |
| Number of Links | 48 |
| Link Bandwidth (ToR Layer) | 1 GB/s |
| Link Bandwidth (Aggregation Layer) | 2 GB/s |
| Link Bandwidth (Core Layer) | 4 GB/s |
| Number of Flows | 300 |
| Flow Bandwidth | 1KB - 200MB |
| Average Packet Rate | 10 pkt/sec |
| Max Packet Size | 1500 Bytes |
| Switch Interface Rate | 1 Gbps |
| MF Size | 1 KB–100 KB |
| EF Size | 101 KB–200 MB |
| MF Size: EF Size | 1:9 |

TABLE VII
AVERAGE ± STANDARD DEVIATION FOR ALL EVALUATED METRICS

| Methods | PLR | MFs FCT | EFs FCT | Goodput | Packet size |
| --- | --- | --- | --- | --- | --- |
| KP-PSO | 2.69 ± 0.72 | 0.18 ± 0.284 | 8.82 ± 2.53 | 457.4 ± 132.04 | 11090.6 ± 3201.58 |
| Sonum | 3.51 ± 0.97 | 0.21 ± 0.282 | 12.02 ± 3.45 | 375.6 ± 108.42 | 9199.2 ± 2655.58 |
| Hedera | 5.05 ± 1.42 | 0.27 ± 0.278 | 14.43 ± 4.15 | 262 ± 75.63 | 7642.6 ± 2206.22 |
| ECMP | 7.09 ± 2.02 | 0.33 ± 0.273 | 19.55 ± 5.63 | 262 ± 75.63 | 6769.8 ± 1954.27 |

**Sonum:** Sonum [13] is a sampling approach for EF detection. It performs detection by consolidating EF information. Then it schedules the preparation of a minimum Packet Loss Rate (PLR) path for EFs.

**ECMP:** ECMP [31] randomly distributes flows among multiple paths by hashing the packet header fields that identify a flow.

**Hedera:** Hedera [53] performs EF detection by polling edge switches for flow statistics in a configurable interval.

### B. Experimental results and discussion

The present study evaluates the Size-KP-PSO approach in terms of the following metrics, which are crucial data center performance parameters occurring in concurrent flows. Table VII presents the average and the standard deviation of all methods and metrics gained through dynamic run of 10 iterations per traffic pattern in each method (50 tests per method).

*1) Packet Loss Rate*

The PLR in DCNs exerts a harmful effect on applications. PLR refers to the failure of packets to reach their destination on a given network. The PLR in a certain interval, can be calculated as: $PLR = \frac{A - A'}{A} * 100\%$ where $A$ and $A'$ are the total number of transmitted and received packets, respectively [54]. This most commonly arises as a consequence of unmanaged network congestion. When the network load exceeds its capacity, it causes the throughput to drop due to bulk transfers. Furthermore, mitigating PLR reduces data retransmission. Fig. 7 illustrates the PLR of four schemes under concurrent flow request traffic loads. Similar to Sonum [13], the link loss rate in the current work is 1% in all schemes, and iperf is utilized to acquire the network PLR. Simulation results show that Size-KP-PSO successfully achieves the lowest PLR in all traffic patterns (avg 2.69%). Indeed, as shown in Fig. 7, Size-KP-PSO presents an average PLR improvement of 1.30x, 1.87x, and 2.63x compared to Sonum, Hedera, and ECMP, respectively.

The main reason for PLR improvement is the present study's two-step flow forwarding model which provides higher bandwidth for each step and also lets the EFs traverse the link. Besides, it prevents MFs from getting stuck between EFs. The PLR reduction also implies an improvement in throughput rate and link utilization. ECMP reports the highest average PLR results (7.09%) and suffers from a lack of dynamic scheduling ability. Hence, ECMP does not take into account bandwidth utilization and the requested flow size of the current network. Both Hedera and Sonum schedule EFs and produce an average PLR of 5.05% and 3.51%, respectively. Hedera performs EF detection through periodic polling from each of the edge-switches and extracts per-flow statistics which lead to switch congestion. Sonum performs sampling to detect EFs and schedules the optimization target to arrange the minimum PLR path for EFs. However, Sonum does not consider the major requirements of MFs and also neglects to perform a fine-grained EF scheduling.

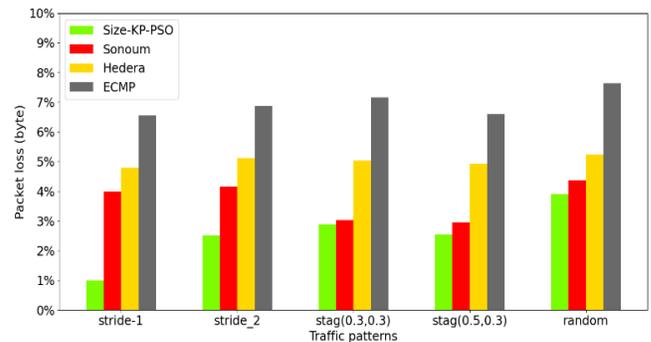
Fig. 7. Overall PLR results in all traffic patterns

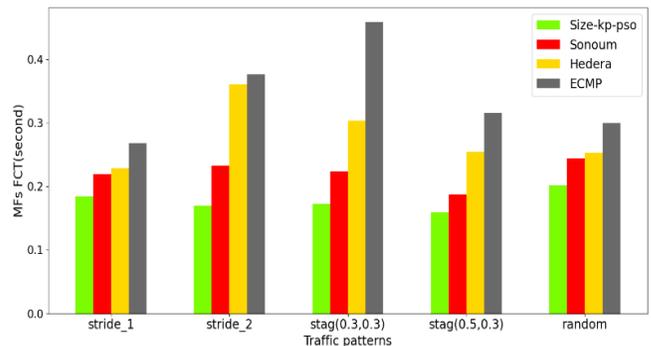
Fig. 8. MFs FCT results in all traffic patterns

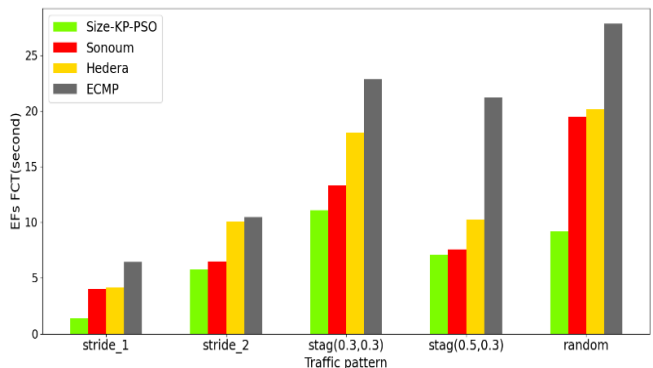
Fig. 9. EFs FCT results in all traffic patterns

In addition, a noteworthy insight into Fig. 7's results are related to the maximum reduction of PLR gained by Size-KP-PSO (0.99%) in stride_1. This is a traffic pattern that expects bottlenecks in some cases due to only a single hop to the destination with the least bandwidth. This improvement points to the significance of the present research's two-step flow forwarding method (Size-KP-PSO) that is perfectly incorporated with the multithreading flow forwarding technique. This also affirms the proposed method's ability to overcome the pressure imposed on the network by concurrent flows.

*2) Flow Completion Time*

FCT is a fundamental metric in the DCN service-level agreement. The FCT is calculated as: $FCT = t_r - t_s$ where $t_r$ and $t_s$ represent the time that the last packet of the flow arrives at the destination server, and the time when the first packet of a flow leaves the source server, respectively. FCT exerts a major impact on delay-sensitive flows (MFs). Moreover, missing FCT deadlines leads to performance degradation. Fig. 8 and 9 present the FCT of MFs and EFs of four schemes, respectively. Size-KP-PSO achieves the highest performance for both MFs (avg 0.18s) and EFs (avg 8.82s) in all traffic patterns.

As Fig. 8 shows, Size-KP-PSO outperforms 1.16x, 1.5x, and 1.83x in terms of MF FCTs as compared with Sonum, Hedera, and ECMP, respectively.

Although Sonum outperforms Hedera and ECMP, it does not consider any specific plan for MFs. Sonum utilizes ECMP only for MFs, but it does not inspect MFs and EFs. Therefore, ECMP's performance is weaker than that of Sonum in MF FCTs. Moreover, Hedera does not focus on MF requirements. ECMP in stag (0.3,0.3) exposes the highest MF FCT. Moreover, as Sonum adopts ECMP for MFs, the ECMP performance is one of the determinative factors in Sonum's MF FCTs alongside other factors (e.g., delays due to MF verification and the adopted controller modules along with the probabilities in stag and random traffic patterns).

Fig. 9 also verifies that, by applying smooth classifications of flow sizes, even the EF FCTs experience major improvement in comparison with other schemes (avg 8.82s shows improvements of 1.36x, 1.63x, and 2.21x as compared with Sonum, Hedera, and ECMP, respectively). Altogether, as EFs constitute 90% of the total DCN traffic, using ECMP for EFs causes a major increase in FCT. ECMP represents the highest EF FCT in all scenarios (avg 19.55s). The Hedera algorithm shows a dynamic slow reaction for flow reallocation while traffic patterns change. This explains why the random traffic pattern results in the highest EF FCT. The Sonum performance relies on the size of the EF set and path set. Hence, the more flows or paths, the larger the search space. Also, the Sonum FCT is limited by the size of the solution space. Hedera and Sonum show an average of 14.43s and 12.02s in EF FCT results, respectively.

All methods represent their highest FCT in the random traffic pattern. Besides, stride_1 reports the lowest EF FCT among all the results.

The results verify that EF detection techniques (e.g, Sonum and Hedera) do not directly consider MFs and are not capable of making distinctions among various EF sizes. Furthermore, applying different approaches for EFs and MFs imposes more overhead on the controller. However, Size-KP-PSO offers a uniform solution that integrates flow size requirements for both MFs and EFs under a saturated workload due to concurrent flow requests.

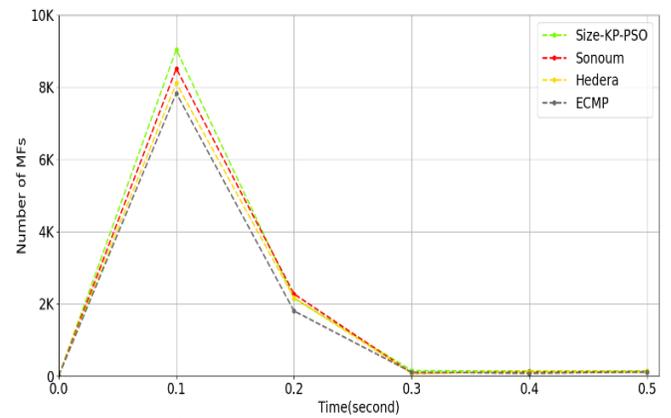

Fig. 10. Number of MFs in initial time (vital time)

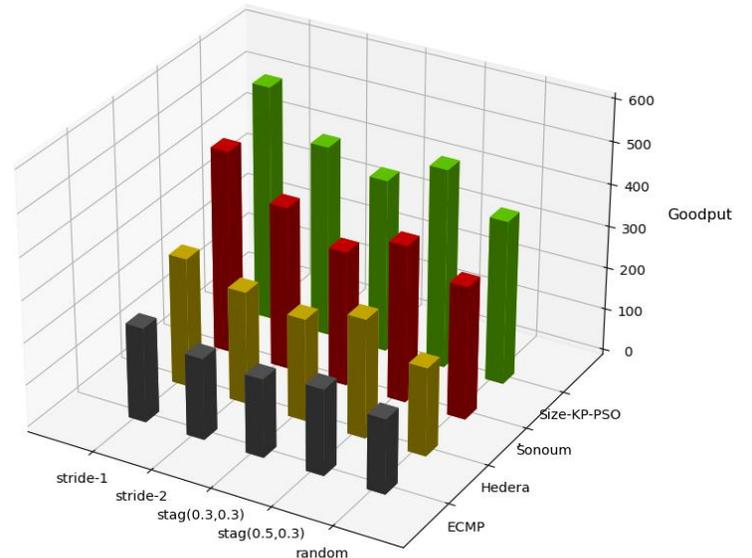

Fig. 11. Average goodput of all methods in each traffic pattern

Furthermore, as MFs are delay-sensitive, Fig. 10 illustrates the number of received MFs (in all performed tests) from the very start (vital time). As Fig. 10 demonstrates, Size-KP-PSO represents the highest number of received MFs through 0.1s, which indicates a 1.06x, 1.11x, and 1.15x improvement compared with Sonum, Hedera, and ECMP, respectively. Since Size-KP-PSO picks up flows with lower sizes (higher values) in selected-results and first forwards them, it is capable of receiving major MFs sooner. Thereafter, Size-KP-PSO maintains a slight and smooth trend in the number of received MFs during the time remaining to the end of receiving the flows of selected-results. Altogether, Size-KP-PSO achieves a 1.04x, 1.10x, and 1.17x higher number of received MFs from the start till 0.5s, which is a determinative time span for MFs when compared with Sonum, Hedera, and ECMP, respectively.

*3) Goodput*

Goodput is defined by only the throughput of the original data. Indeed, goodput is equal to throughput without traffic overhead (packet headers and retransmitted data) and it is calculated as: $goodput = \frac{max_r}{total_s}$ where $max_r$ and $total_s$ are the maximum number of packets received by the receiver in sequence, and the total number of packets sent by the sender including retransmissions, respectively [55]. Since EFs are large in size, they consume more bandwidth and are, in essence, throughput-sensitive. Hence, a higher goodput reflects a higher and pure amount of throughput, which results in improving EF satisfaction.

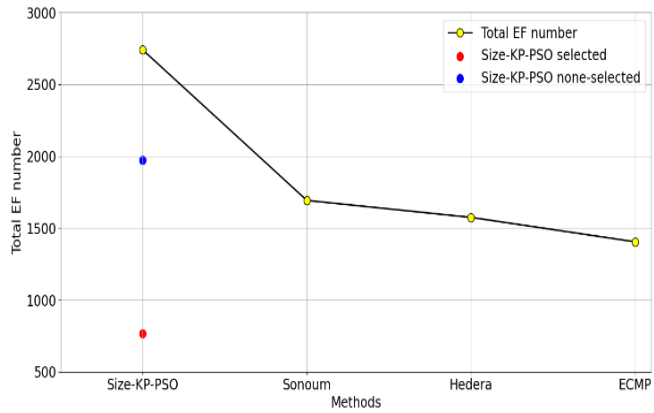

Fig. 12. Total EF number of all methods in each traffic pattern

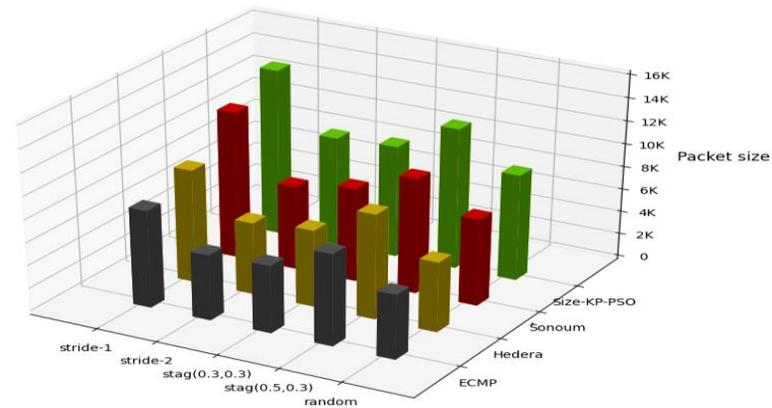

Fig. 13. The packet size of all methods in each traffic pattern

Fig. 11 indicates the average goodput results in all traffic patterns for each method. Size-KP-PSO gains 1.21x, 1.74x, and 2.31x higher goodput when compared with Sonum, Hedera, and ECMP, respectively. In Size-KP-PSO, EFs are marked with values (priority) according to their size (the smaller size, the higher the value). Therefore, as Fig.11 demonstrates, the combination of EFs forwarded in selected-results include a minor population of EFs with lower sizes. Also, the combination of EFs forwarded in non-selected-results make up the major population of EFs with higher sizes. As Fig. 12 depicts, Size-KP-PSO achieves improvement of 1.61x, 1.73, and 1.95x in the number of EFs as compared with Sonum, Hedera, and ECMP, respectively.

Furthermore, as Fig. 13 illustrates, Size-KP-PSO transfers 1.20x, 1.45x, and 1.61 packet sizes in comparison with Sonum, Hedera, and ECMP, respectively. Since packet size is to some extent proportional to flow size, this also reconfirms that the proposed method succeeds in preparing better conditions for EFs to traverse the link while respecting MFs.

## V. CONCLUSION

The current research introduces a flow-aware method (Size-KP-PSO) for scheduling flows in SDN-based data center networks. The proposed approach distinguishes flow priorities by their ToS values. The present study models the problem as a knapsack model and dedicates the ToS tag and flow volume to value and weight vectors, respectively. Finally, to obtain the appropriate selected-results vector, the problem is solved with the PSO algorithm. The results show that the current study succeeds in improving PLR, the FCT of MFs and EFs, and the number of received MFs in vital time. Moreover, the proposed method achieves higher goodput and average delivered packet size when compared to Sonum, Hedera, and ECMP schemes. This, in turn, leads to significant quality of service improvement.